\documentclass[aps,showpacs,nofootinbib]{revtex4}

\usepackage{graphicx}
\usepackage{amsmath}
\usepackage{amssymb}
\usepackage{bm}

\newcommand{\be}{\begin{equation}}
\newcommand{\ee}{\end{equation}}
\newcommand{\ba}{\begin{eqnarray}}
\newcommand{\ea}{\end{eqnarray}}
\newcommand{\tr}{\mbox{tr}}

\def\lb{\label}

\def\fsl#1{\setbox0=\hbox{$#1$}                 
   \dimen0=\wd0                                 
   \setbox1=\hbox{/} \dimen1=\wd1               
   \ifdim\dimen0>\dimen1                        
      \rlap{\hbox to \dimen0{\hfil/\hfil}}      
      #1                                        
   \else                                        
      \rlap{\hbox to \dimen1{\hfil$#1$\hfil}}   
      /                                         
   \fi}                                         %

\begin{document}

\title{Nondecoupling phenomena in QED in a magnetic field and
noncommutative QED}

\author{E.V. Gorbar}
  \email{egorbar@uwo.ca}
  \altaffiliation[On leave from ]{
       Bogolyubov Institute for Theoretical Physics,
       03143, Kiev, Ukraine}

\author{Michio Hashimoto}
  \email{mhashimo@uwo.ca}

\author{V.A. Miransky}
  \email{vmiransk@uwo.ca}
   \altaffiliation[On leave from ]{
       Bogolyubov Institute for Theoretical Physics,
       03143, Kiev, Ukraine
}
   
\affiliation{
Department of Applied Mathematics, University of Western
Ontario, London, Ontario N6A 5B7, Canada
}

\date{\today}

\begin{abstract}
The dynamics in QED in a strong constant magnetic field 
and its connection with the noncommutative QED are studied.
It is shown that in the regime with the lowest Landau
level (LLL) dominance the $U(1)$ gauge symmetry
in the fermion determinant is transformed
into the noncommutative $U(1)_{nc}$ gauge symmetry. In this regime,
the effective action is intimately connected with that in 
noncommutative QED and the original $U(1)$ gauge Ward identities are 
broken (the LLL anomaly). On the other hand, it is 
shown that although a contribution
of each of an infinite number of higher Landau levels is suppressed in an
infrared region, their cumulative contribution is not 
(a nondecoupling
phenomenon). This leads to a restoration of the original $U(1)$ gauge 
symmetry in the infrared dynamics. The physics underlying this phenomenon 
reflects the important role of a boundary dynamics at spatial infinity
in this problem. 

\end{abstract}


\maketitle

Since the classical papers \cite{HEW,Sch}, 
the problem of QED in a constant magnetic 
field has been thoroughly studied (for a recent review,
see Ref. \cite{D}). 
In this letter, we consider
this 
problem for the case of a strong magnetic field. In particular, we
study the connection of this dynamics with that in noncommutative 
QED (for reviews of noncommutative field theories (NCFT), see
Ref. \cite{DNS}). The motivation for this study was the recent results 
obtained in
Ref. \cite{GM}, where the connection between the dynamics in
the Nambu-Jona-Lasinio (NJL) model
in a strong magnetic field
and that in NCFT was established.
The main conclusion of that paper was
that although in the regime with the
lowest Landau level (LLL) dominance the NJL model determines a 
NCFT, this NCFT is different from
the conventional ones considered in the literature.
In particular,
the UV/IR mixing, taking place in the
conventional NCFT \cite{MRS}, is absent in this case.
The reason of that
is an inner structure (i.e., dynamical form-factors) of neutral
composites in this model.

In this letter, some sophisticated features of the dynamics in QED
in a strong magnetic field are revealed. It is shown that
in the approximation with the 
lowest Landau
level (LLL) dominance, the initial $U(1)$ gauge symmetry
in the fermion determinant is transformed
into the noncommutative $U(1)_{nc}$ gauge symmetry. In this regime,
the effective action is intimately connected with that in
noncommutative QED and the original $U(1)$ gauge Ward identities are
broken (we call this phenomenon an LLL anomaly). In fact, this
dynamics yields a modified noncommutative QED
in which the UV/IR mixing is absent,
similarly to the case of the NJL model in a strong magnetic field
\cite{GM}. The reason of that
is an inner structure (i.e., a dynamical form-factor) of photons
in a strong magnetic field. However, it is not the end of the story.
We show that adding the
contribution of {\it all} the higher Landau levels removes the LLL
anomaly and restores the original $U(1)$ gauge symmetry. This
restoration happens in a quite sophisticated way: although 
a contribution
of each of an infinite number of higher Landau levels is suppressed in an
infrared region, their cumulative contribution is not
(a nondecoupling phenomenon). As will be discussed below, this  
phenomenon reflects the important role of a boundary dynamics 
at spatial infinity
in this problem. We also indicate the kinematic region where the
LLL approximation is reliable.

To put the dynamics in QED in a magnetic field under control, we
will consider the case with a large number of fermion flavors $N$,
when the $1/N$ expansion is reliable. We also choose the 
current fermion mass $m$ satisfying the condition
$m_{dyn}\ll m \ll \sqrt{|eB|}$,
where $m_{dyn}$ is the dynamical mass of fermions generated
in the chiral symmetric QED in a magnetic 
field \cite{QED1}. 
\footnote {The dynamical mass is
$m_{dyn} \simeq \sqrt{|eB|} \exp\left(-N\right)$
for a large running coupling $\tilde{\alpha}_{b} \equiv
N\alpha_{b}$
related to the magnetic scale $\sqrt{|eB|}$, and
$m_{dyn} \sim \sqrt{|eB|}\exp\left[-\frac{\pi N}
{\tilde{\alpha}_{b}\ln\left(1/\tilde{\alpha}_{b}\right)}\right]$
when the coupling $\tilde{\alpha}_{b}$ is weak \cite{QED1}.}
The condition $m_{dyn}\ll m$ guarantees that there are no light
(pseudo) Nambu-Goldstone bosons, and the only particles in the low energy 
effective 
theory in this model are photons. 
As to the condition  $m \ll 
\sqrt{|eB|}$,
it implies that the magnetic field is very strong.

Integrating out fermions, we obtain
the effective action for photons in the leading order in $1/N$:
\begin{equation}
 \Gamma = \Gamma^{(0)} + \Gamma^{(1)}, \quad
 \Gamma^{(0)} = - \frac{1}{4}\int d^4x \,f_{\mu\nu}^2, \quad 
 \Gamma^{(1)} =
 -iN\mbox{Tr}\mbox{Ln} \left[ i\gamma^{\mu} (\partial_{\mu} 
 -ieA_{\mu}) -m\, \right],
 \lb{ACTION}
\end{equation}
where $f^{\mu\nu} = \partial_{\mu}A_{\nu} - \partial_{\nu}A_{\mu}$
and the vector field $A_{\mu} = A^{cl}_{\mu} + \tilde{A}_{\mu}$,
where the classical part $A^{cl}_{\mu}$ is $A^{cl}_{\mu} =
\langle 0|A_{\mu}|0 \rangle$. Since the constant magnetic field $B$ is a
solution of the exact equations in QED (see for example the discussion in 
Sec. 2
in the second paper in Ref. \cite{QED1}), it is   
\ba
A_{\mu}^{cl} = (0,\frac{Bx^2}{2},-\frac{Bx^1}{2},0).
\label{symm}
\ea
This field describes a constant magnetic field directed in the $+x^3$
direction and we use the so called symmetric gauge for $A_{\mu}^{cl}$.

For a strong magnetic field
$|eB| \gg m^2$, it is naturally to expect that in the infrared region 
with 
momenta $k \ll \sqrt{|eB|}$, the LLL approximation should be
reliable. Indeed, let us consider the fermion propagator
in a magnetic field \cite{Sch}:
\begin{eqnarray}
S(x,y)=\exp\biggl[\frac{ie}{2}(x-y)^\mu A_\mu^{\rm ext}(x+y)\biggr]
\tilde S(x-y)~, 
\label{propagator}
\end{eqnarray}
where the Fourier transform of the translationally 
invariant part $\tilde S$ can be decomposed over the Landau levels
\cite{Chod}:
\ba
\tilde S(k)=i \exp\biggl(-\frac{{\bm k}_{\perp}^2}{|eB|}\biggr)
\sum^\infty_{n=0}
(-1)^n \frac{D_n(eB,k)}{k_{\|}^{2}-m^2-2|eB|n}
\lb{decomposition}
\ea
with ${\bm k}_{\perp}\equiv (k^1,k^2)$
and $k_{\|} \equiv (k_0, k_3)$. 
The
functions $D_n(eB,k)$ are expressed through the
generalized Laguerre polynomials $L_m^\alpha$:
\begin{align}
D_n(eB,k) & \,=\, (k_{\|}\gamma^{\|} +m)
\biggl[\bigl(1-i\gamma^1\gamma^2 {\rm sign}(eB)\bigr)L_n
\biggl(2\frac{{\bm k}_{\perp}^2}{|eB|}\biggr)
-\bigl(1+i\gamma^1\gamma^2{\rm sign}(eB)\bigr)L_{n-1}
\biggl(2\frac{{\bm k}_{\perp}^2}{|eB|}\biggr)\biggr]
 \nonumber\\ & \qquad 
+4(k^1 \gamma^1+k^2\gamma^2)L^1_{n-1}
\biggl(2\frac{{\bm k}_{\perp}^2}{|eB|}\biggr)\, ,
\lb{L}
\end{align}
where $\gamma^{\|} \equiv (\gamma^0,\gamma^3)$.
Relation (\ref{decomposition}) seems to 
suggest that in the infrared
region, with $k_{\perp}, k_{\|} \ll \sqrt{|eB|}$, 
all the higher Landau levels with
$n \geq 1$ decouple and only the LLL with $n = 0$ is relevant.
Although this argument is physically convincing, there may be a
potential flaw due to an infinite number of the Landau levels. As
will be shown below, this is indeed the case in this problem:
the cumulative contribution of the higher Landau levels does not
decouple. 

But first we will consider the dynamics in the LLL
approximation.
In this case, the calculation of the effective action
(\ref{ACTION}) 
is reduced to calculating fermion loops with the LLL fermion 
propagators. Such a problem in the 
NJL model in a magnetic field has been
recently solved in Ref. \cite{GM}. The extension of that
analysis to the case of QED is straightforward. The effective
action (\ref{ACTION}) in the LLL approximation is given by
\begin{equation}
 \Gamma_{LLL} = \Gamma^{(0)} + \Gamma^{(1)}_{LLL}, \quad  
 \Gamma^{(1)}_{LLL} = -\frac{iN|eB|}{2\pi} \int d^2x_{\perp}\,
 \mbox{Tr}_{||}\Big[{\cal P}\,
 \mbox{Ln}[i\gamma^{||}(\partial_{||} - ie{\cal A}_{||})-m]\Big]_{*} 
 \lb{starACTION}
\end{equation}
(compare with Eq. (54) in \cite{GM}). Here $*$ is the 
symbol of 
the Moyal star product, which is a signature of a NCFT
\cite{DNS}, the projector ${\cal P}$ is
${\cal P} \equiv [1 - i\gamma^1\gamma^2 \mbox{sign}(eB)]/2$,
and the longitudinal ``smeared" fields 
${\cal A}_{||}$ are defined as 
${\cal A}_{||} =
e^{\frac{\nabla_{\perp}^2}{4|eB|}} A_{||}$ \cite{GM}, 
where $\nabla_{\perp}^2$ is the transverse Laplacian.
Notice that ${\cal P}$ is the
projector on the fermion (antifermion) states with the spin
polarized along (opposite to) the magnetic field
and that the one-loop term $\Gamma^{(1)}_{LLL}$ in
(\ref{starACTION}) includes only the longitudinal field
${\cal A}_{||}  =
({\cal A}_{0}, {\cal A}_{3})$. This is because
the LLL fermions couple only to the longitudinal components of
the photon field \cite{QED1}.

In action (\ref{starACTION}), the trace $\mbox{Tr}_{||}$, related to the 
longitudinal subspace, is taken in the functional sense and the
star product 
relates to the space transverse coordinates. Therefore
the LLL dynamics 
determines a NCFT with noncommutative
transverse coordinates $\hat{x}^{a}_{\perp}$, $a = 1,2$:
\ba
[\hat{x}^{a}_{\perp}, \hat{x}^{b}_{\perp}] = 
i\frac{1}{eB}\epsilon^{ab} \equiv i\theta^{ab}.
\label{commrel}
\ea

The structure of the logarithm of the fermion determinant
in $\Gamma^{(1)}_{LLL}$ (\ref{starACTION}) implies that it is invariant 
not under
the initial $U(1)$ gauge symmetry but under the noncommutative
$U(1)_{nc}$ gauge one \cite{DNS} (henceforth we omit the subscript $||$ in
gauge fields):   
\begin{subequations}
\begin{align}
{\cal A}_{\mu} &\;\to\; U(x)*{\cal A}_{\mu}*U^{-1}(x) +
\frac{i}{e}U(x)*\partial_{\mu}U^{-1}(x), && (\mu=0,3)  \\[2mm]
{\cal F}_{\mu\nu} &\;\to\; U(x)*{\cal F}_{\mu\nu}*U^{-1}(x),
&& (\mu,\nu=0,3) 
\label{gt}
\end{align}
\end{subequations}
where $U(x)=(e^{i\lambda(x)})_{*}$ and the field strength 
${\cal F}_{\mu\nu}$ is
\begin{equation}
{\cal F}_{\mu\nu} = 
\partial_{\mu}{\cal A}_{\nu} - \partial_{\nu}{\cal A}_{\mu}
-ie[{\cal A}_{\mu},{\cal A}_{\nu}]_{\rm MB}^{}
\label{F}  
\end{equation}
with the Moyal bracket 
\begin{equation}
[{\cal A}_{\mu},{\cal A}_{\nu}]_{\rm MB}^{} \equiv 
{\cal A}_{\mu}*{\cal A}_{\nu} - {\cal A}_{\nu}*{\cal A}_{\mu}.   
\end{equation}

Therefore the derivative expansion of 
$\Gamma^{(1)}_{LLL}$ should be
expressed through terms with the star product of the field ${\cal 
F}_{\mu\nu}$ 
and its covariant derivatives:
\begin{equation}
  \Gamma^{(1)}_{LLL} = 
   a_0 S_{{\cal F}^2}^{} + a_1 S_{{\cal F}^3}^{}
 + a_2 S_{({\cal D}{\cal F})^2}^{}
 + a_3 S_{{\cal D}^2 {\cal F}^2}^{} + \cdots \,\,\,,
\lb{expansion}
\end{equation}
where
\begin{align}
  S_{{\cal F}^2}^{} &\equiv 
 -\frac{1}{4}\int d^2 x_{\perp}^{} d^2 x_{\parallel}^{}\,
 {\cal F}_{\mu\nu}*{\cal F}^{\mu\nu}, \quad &
  S_{{\cal F}^3}^{} &\equiv ie \int d^2 x_{\perp}^{} d^2 x_{\parallel}^{}\,
 {\cal F}_{\mu\nu} * {\cal F}^{\nu\lambda} * {\cal F}_\lambda^{\;\;\mu},
 \nonumber \\
  S_{({\cal D} {\cal F})^2}^{} &\equiv \int d^2 x_{\perp}^{} 
  d^2 x_{\parallel}^{}\,
 {\cal D}_\lambda {\cal F}^{\lambda\mu} * 
 {\cal D}^\rho {\cal F}_{\rho\mu}, \quad &
  S_{{\cal D}^2 {\cal F}^2}^{} &\equiv \int d^2 x_{\perp}^{} 
  d^2 x_{\parallel}^{}\,
 {\cal D}_\lambda {\cal F}_{\mu\nu} * 
 {\cal D}^\lambda {\cal F}^{\mu\nu}, \quad 
\end{align}
and the covariant derivative of ${\cal F}_{\mu\nu}$ is
${\cal D}_\lambda {\cal F}_{\mu\nu} = 
  \partial_{\lambda}{\cal F}_{\mu\nu} 
-ie[{\cal A}_{\lambda},{\cal F}_{\mu\nu}]_{\rm MB}^{}$.
These are all independent operators which have the dimension four
and six. In particular, by using the Jacobi identity,
\begin{equation}
 [{\cal D}_\mu,[{\cal D}_\nu,{\cal D}_\lambda]_{\rm MB}^{}]_{\rm MB}^{} + 
 [{\cal D}_\nu,[{\cal D}_\lambda,{\cal D}_\mu]_{\rm MB}^{}]_{\rm MB}^{} +
 [{\cal D}_\lambda,[{\cal D}_\mu,{\cal D}_\nu]_{\rm MB}^{}]_{\rm MB}^{} = 0,
\end{equation}
and the relation
${\cal F}_{\mu\nu}=ie^{-1}[{\cal D}_\mu,{\cal D}_\nu]_{\rm MB}^{}$, 
one can easily check that the operator
$\int d^2 x_{\perp}^{}d^2 x_{\parallel}^{}\,
 {\cal D}_\lambda {\cal F}_{\mu\nu} * {\cal D}^\mu {\cal F}^{\nu\lambda}$ 
is not independent: it is equal to $-1/2\,S_{{\cal D}^2 {\cal F}^2}^{}$.

The coefficients $a_i$, $(i=0,1,2,3,\cdots)$ 
in Eq. (\ref{expansion}) can be found from the
$n$-point photon vertices 
\begin{equation}
T_{LLL}^{(n)} = i\frac{(ie)^{n} N|eB|}{2\pi n} \int d^2x^{\perp}\,
d^2x^{||}_1 \cdots d^2x^{||}_n\,
\,\,\mbox{tr}
\left[S_{||}(x^{||}_1-x^{||}_2)
{\hat{{\cal A}}}(x^{\perp},x^{||}_2)\,...\,
S_{||}(x^{||}_n-x^{||}_1)
{\hat{{\cal A}}}(x^{\perp},x^{||}_1)\right]_{*}
\label{nvertex}
\end{equation}
by expanding the vertices in powers of external momenta
(here $S_{\|}(x_{\|}) = \int \frac{d^2k_{\|}}{(2\pi)^2}
e^{-ik_{\|}x^{\|}} \frac{i}{k_{\|}\gamma^{\|} - m}\,
{\cal P}$ and 
${\hat{{\cal A}}} \equiv \gamma^{||}{\cal A}_{||}$).
In particular, from 
the vertices $T_{LLL}^{(2)}$ and $T_{LLL}^{(3)}$, we
find
the coefficients $a_0$, $a_1$, $a_2$, and $a_3$ connected with
the operators of the dimension four and six in the 
derivative expansion (\ref{expansion}) of $\Gamma^{(1)}_{LLL}$: 
\ba
 a_0 = \frac{\tilde{\alpha}}{3\pi}\frac{|eB|}{m^2}, \quad
 a_1 = \frac{1}{60m^2}a_0, \quad a_2 = -\frac{1}{10m^2}a_0, \quad a_3=0,
\lb{coeff}
\ea
where $\tilde{\alpha} \equiv N\alpha = Ne^2/(4\pi)$
(since in the presence of
a magnetic field the charge conjugation symmetry is broken,
the 3-point vertex is nonzero).

The $U(1)$ gauge Ward identities imply that the $n$-point photon amplitude
$T^{\mu_{1}...\mu_{n}}(x_{1},...,x_{n})$ should be
transverse, i.e.,
$\partial_{\mu_i} T^{\mu_{1}...\mu_{n}}(x_{1},...,x_{n}) = 0$.
It is easy to show that the 2-point vertex 
$T_{LLL}^{\mu_1\mu_2}$
yielding the
polarization operator is transverse indeed. 
Now let us turn to the 3-point vertex and show that it is not
transverse, i.e., the Ward identities connected with the initial
gauge $U(1)$ are broken in the LLL approximation. 
In the momentum space, the vertex is
\begin{equation}
  T_{LLL}^{\mu_1\mu_2\mu_3}(k_1,k_2,k_3) = Ne^3 \frac{|eB|}{2\pi}
  \sin\left(\frac{1}{2}\theta_{ab}k_{1\,\perp}^a k_{2\,\perp}^b\right)
  \Delta_{LLL}^{\mu_{1\parallel}\mu_{2\parallel}\mu_{3\parallel}}
  (k_{1\,\parallel},k_{2\,\parallel},k_{3\,\parallel})
\lb{3-point}
\end{equation}
with
\footnote{
The explicit expression for 
$\Delta_{LLL}^{\mu_{1\parallel}\mu_{2\parallel}\mu_{3\parallel}}$
is given in terms of
the two dimensional version of the Passarino-Veltman
functions \cite{Passarino}. It is quite cumbersome and will
be written down elsewhere.}
\begin{equation}
 \Delta_{LLL}^{\mu_{1\parallel}\mu_{2\parallel}\mu_{3\parallel}}
 (k_{1\,\parallel},k_{2\,\parallel},k_{3\,\parallel})
 \equiv \int \frac{d^2 \ell_\parallel}{i(2\pi)^2}
 \frac{\tr\Big[\gamma^{\mu_1}_\parallel\,[(\fsl{\ell}-\fsl{k}_1)_\parallel+m]\,
               \gamma^{\mu_2}_\parallel\,[(\fsl{\ell}+\fsl{k}_3)_\parallel+m]\, 
               \gamma^{\mu_3}_\parallel\,(\fsl{\ell}_\parallel+m)\Big]}
      {(\ell_\parallel^2-m^2)[(\ell-k_1)_\parallel^2-m^2]
       [(\ell+k_3)_\parallel^2-m^2]}.
\end{equation}
The argument of the sine in Eq. (\ref{3-point}) is the Moyal cross product 
with 
$\theta_{ab}
= \epsilon^{ab}/eB$ (see Eq. (\ref{commrel})).
It is easy to find that the transverse part of 
the vertex (\ref{3-point}) is not zero and equal to
\ba
k_{1\mu_1} T_{LLL}^{\mu_1\mu_2\mu_3}(k_1,k_2,k_3) =
-\frac{2e}{i}\,
\sin\left(\frac{1}{2}\theta_{ab}k_{1\,\perp}^a k_{2\,\perp}^b\right)
\Big[\,\Pi_{\,\parallel}^{\mu_2\mu_3}(k_{2\,\parallel}) - 
       \Pi_{\,\parallel}^{\mu_2\mu_3}(k_{3\,\parallel})\,\Big],
\label{divergence}
\ea
with
\ba
\Pi_{\,\parallel}^{\mu\nu}(k_{\parallel}) = i\frac{2\tilde{\alpha}|eB|}{\pi}
\left(\,g_{\,\parallel}^{\mu\nu} - 
\dfrac{k_{\parallel}^{\mu}k_{\parallel}^{\nu}}
{k_{\parallel}^2}\right) \Pi(k_{\parallel}^2), \qquad
\Pi(k_{\parallel}^2) \equiv 
1 + \dfrac{2m^2}{k_{\parallel} ^2\sqrt{1-\frac{4m^2}
{k_{\parallel}^2}}}
        \ln\frac{\phantom{-}1 + \sqrt{1-\dfrac{4m^2}
{k_{\parallel}^2}}}
                {-1+\sqrt{1-\dfrac{4m^2}
{k_{\parallel}^2}}} \; .
\label{polarization}
\ea
Here we defined $g_{\parallel}^{}=\mbox{diag}(1,-1)$
[note that $\Pi_{\,\parallel}^{\mu\nu}$ coincides with the polarization tensor
in the (1+1)-dimensional QED if the parameter $2\tilde{\alpha}|eB|$
here is replaced by the coupling $e^{2}_1$ in $\mbox{QED}_{1+1}$].
Thus, the original $U(1)$ gauge Ward identities are broken in the
LLL approximation. We will call this an LLL anomaly.

Note that all the vertices $T^{(n)}_{LLL}$ with $n \geq 3$ are
finite and a logarithmic divergence in the vertex $T^{(2)}_{LLL}$,
which is proportional to the polarization operator, is absent if
one uses a gauge invariant regularization. In fact, it is
sufficient if the regularization is invariant under the
longitudinal $U(1)_{||}$ gauge group with phase parameters
$\alpha(x^{||})$
depending only on longitudinal coordinates. This $U(1)_{||}$
is a subgroup of both the gauge $U(1)$ and the noncommutative
gauge $U(1)_{nc}$ and it is the gauge symmetry of the whole
action $\Gamma_{LLL}$ (\ref{starACTION}). Indeed, while 
the free Maxwell term $\Gamma^{(0)}$ in (\ref{starACTION}) is 
invariant under the gauge $U(1)$, 
the one-loop term $\Gamma^{(1)}_{LLL}$ is
invariant under the $U(1)_{nc}$. Another noticeable point 
is that the divergence
(\ref{divergence}) is not a polynomial function of momenta
and have branch point singularities. Therefore
the fact that $T^{(3)}_{LLL}$ is not transverse is
regularization independent. 

The origin of the LLL anomaly is clear:
the $T^{(n)}_{LLL}$ vertices come from the one-loop part
$\Gamma^{(1)}_{LLL}$ of the action which is invariant not
under the  gauge $U(1)$ but under the $U(1)_{nc}$. Therefore the
Ward identities for the vertices $T^{(n)}_{LLL}$ reflect
not the $U(1)$ gauge symmetry but the noncommutative symmetry
$U(1)_{nc}$. 

Notice that
the action $\Gamma_{LLL}$ (\ref{starACTION}) determines a
conventional noncommutative QED only in the case of an induced 
photon field, when the Maxwell term $\Gamma^{(0)}$ is absent.
When this term is present, the action also determines a NCFT,
however, this NCFT is different from the conventional ones considered in 
the literature. In particular, expressing the photon field $A_{\mu}$
through the smeared field ${\cal A}_{\mu}$ as
$A_{\mu} = e^{\frac{-\nabla_{\perp}^2}{4|eB|}} {\cal A_{\mu}}$, we
find that the propagator of the smeared field rapidly, as 
$e^{\frac{-p^{2}_{\perp}}{2|eB|}}$, 
decreases for large transverse momenta.
The form-factor $e^{\frac{-p^{2}_{\perp}}{2|eB|}}$
built in the smeared field reflects an inner structure of
photons in a magnetic field.
This feature leads to removing the UV/IR mixing in this NCFT
(compare with the analysis of the UV/IR mixing in Sec. 4 of Ref. 
\cite{GM}). 

It is clear however that, since the initial QED has
the usual $U(1)$ gauge symmetry, there should exist an additional
contribution that restores the $U(1)$ Ward identities broken in the
LLL approximation. We will show that this contribution comes
from heavy (naively decoupled) higher Landau levels (HLL),
and it is necessary to consider the contribution of all
of them in order to restore the Ward identities. 

Let us consider the 3-point vertex replacing one of the LLL propagator
by the full one in Eq.~(\ref{decomposition}). 
We find that at each $n \geq 1$ the contributions in the vertex of
the first, second and third terms in Eq.~(\ref{L}) are
respectively 
\begin{eqnarray}
  \Delta_{HLL}^{\mu_{\parallel}\nu_{\parallel}\lambda_{\parallel}}(p,q,k) &\sim& 
   \frac{(-1)^n}{n|eB|}\; 
   f^{\mu\nu\lambda}_\parallel (p_\parallel,q_\parallel,k_\parallel),
  \lb{one} \\
  \Delta_{HLL}^{\mu_{\perp}\nu_{\perp}\lambda_{\parallel}}(p,q,k) &\sim& 
   \frac{(-1)^n}{n|eB|}\Big[\,\,
   g_{\,\perp}^{\mu\nu} h_1(p_\perp,q_\perp,k_\perp)
 +\epsilon_{\,\perp}^{\mu\nu} h_2(p_\perp,q_\perp,k_\perp)
          \mbox{sign}(eB)\,\Big] \,
  \left(\,q_{\,\parallel}^{\lambda} - 
  \dfrac{(q \cdot k)_{\parallel}}{k_{\parallel}^2}\,k_{\parallel}^{\lambda}\,\right)
  \Pi(k_{\parallel}^2),
  \lb{two} \\
  \Delta_{HLL}^{\mu_{\perp}\nu_{\parallel}\lambda_{\parallel}}(p,q,k) &\sim& 
   \frac{(-1)^n}{n|eB|}\; h^{\mu_\perp}_3 (p_\perp,q_\perp,k_\perp)
   \left(\,g_{\,\parallel}^{\nu\lambda} -
   \dfrac{k_{\parallel}^{\nu}k_{\parallel}^{\lambda}}
   {k_{\parallel}^2}\right) \Pi(k_{\parallel}^2),
  \lb{three}
\end{eqnarray}
where $g_\perp^{}=\mbox{diag}(-1,-1)$,
$(p \cdot q)_\parallel^{}=p^0 q^0 - p^3 q^3$, and
$f^{\mu\nu\lambda}_\parallel$, $h_{1,2}$ and 
$h^{\mu_\perp}_3$ are some smooth functions of 
longitudinal and transverse momenta.
As was expected, the contribution of each of the 
individual HLL with $n \geq 1$ is suppressed by powers of
$1/|eB|$ in the infrared region.
It is however quite remarkable that despite the suppression of individual
HLL contributions, 
their cumulative contribution
is not suppressed in the infrared region (a nondecoupling effect).
In fact, using the relation \cite {GR}
\begin{equation}
(1-z)^{-(\alpha+1)}\exp\biggl(\frac{xz}{z-1}\biggr)=\sum^\infty_{n=0}
L^\alpha_n(x)z^n~
\end{equation}
and integrating it with respect to $z$, 
we can perform explicitly the summation over the HLL contributions and
obtain the 3-point vertex that satisfies the Ward identities
for the $U(1)$ gauge symmetry:
\begin{equation}
T^{\mu_1\mu_2\mu_3} (k_1,k_2,k_3) = 
  T^{\mu_1\mu_2\mu_3}_{LLL}(k_1,k_2,k_3)
+ T^{\mu_1\mu_2\mu_3}_{HLL1}(k_1,k_2,k_3)
+ T^{\mu_1\mu_2\mu_3}_{HLL2}(k_1,k_2,k_3),
\lb{proper}
\end{equation}
where
\begin{eqnarray}
\lb{a}
\lefteqn{
T^{\mu_1\mu_2\mu_3}_{HLL1}(k_1,k_2,k_3) = \frac{2e}{i}\,
  \sin\left(\frac{1}{2}\theta_{ab}k_{1\,\perp}^a k_{2\,\perp}^b\right)
} \nonumber \\ && \times 
\Bigg[\,\dfrac{-k_{1\perp}^{\mu_1}}{{\bm k}_{1\perp}^2}\,
        \left(\,\phantom{\frac{}{}}
                \Pi_{\,\parallel}^{\mu_2\mu_3}(k_{2\parallel})
              - \Pi_{\,\parallel}^{\mu_2\mu_3}(k_{3\parallel})\,\right) 
      + \frac{k_{1\perp}^{\mu_1}k_{2\perp}^{\mu_2}
             -k_{1\perp}^{\mu_2}k_{2\perp}^{\mu_1}
            -({\bm k}_1 \cdot {\bm k}_2)_\perp^{} g_{\,\perp}^{\mu_1\mu_2}}
             {{\bm k}_{1\perp}^2\,{\bm k}_{2\perp}^2}\,\, 
        k_{2\parallel\,\nu}\Pi_{\,\parallel}^{\mu_3\nu}(k_{3\parallel})
\nonumber \\ && \qquad + \,\,
\mbox{permutations of}\,\, (k_1,\mu_1),\, (k_2,\mu_2),\,\,\mbox{and}\,\,
(k_3,\mu_3)\,\Bigg]\,,
\end{eqnarray}
\begin{eqnarray}
\lb{b}
\lefteqn{
T^{\mu_1\mu_2\mu_3}_{HLL2}(k_1,k_2,k_3) = -\frac{2e}{i}\mbox{sign}(eB)
} \nonumber \\ && \times 
\Bigg[\,\left\{\,\exp\left(\,-\frac{({\bm k}_1\cdot{\bm k}_2)_\perp^{}}
                                      {2|eB|}\,\right)
        -\cos\left(\frac{1}{2}\theta_{ab}k_{1\,\perp}^a k_{2\,\perp}^b\right)
        \,\right\} 
\left(\,\frac{k_{1\perp}^{\mu_2}\epsilon_{\,\perp}^{ab}k_{1\,\perp}^a k_{2\,\perp}^b
              +({\bm k}_1 \cdot {\bm k}_2)_\perp^{} \epsilon_{\,\perp}^{\mu_2 b}
               k_{1\,\perp}^b}
              {{\bm k}_{1\perp}^2\,{\bm k}_{2\perp}^2}\,\, 
          \Pi_{\,\parallel}^{\mu_3\mu_1}(k_{3\parallel})
\right. \nonumber \\ && \qquad \left.
        +\frac{k_{2\perp}^{\mu_1}\epsilon_{\,\perp}^{ab}k_{2\perp}^a k_{1\perp}^b
              +({\bm k}_2 \cdot {\bm k}_1)_\perp^{}
              \epsilon_{\,\perp}^{\mu_1 b}k_{2\perp}^b} 
              {{\bm k}_{1\perp}^2\,{\bm k}_{2\perp}^2}\,\, 
          \Pi_{\,\parallel}^{\mu_2\mu_3}(k_{3\parallel})  
        +\frac{g_{\,\perp}^{\mu_1\mu_2}
               \epsilon_{\,\perp}^{ab}k_{1\perp}^a k_{2\perp}^b
              +\epsilon_{\,\perp}^{\mu_1 \mu_2}({\bm k}_1 \cdot {\bm k}_2)_\perp^{}}
              {{\bm k}_{1\perp}^2\,{\bm k}_{2\perp}^2}\,\,\,
          k_{2\parallel\nu} \Pi_{\,\parallel}^{\mu_3\nu}(k_{3\parallel})\,\, \right)
\nonumber\\ && \qquad + \,\, 
\mbox{permutations of}\,\, (k_1,\mu_1),\, (k_2,\mu_2),\,\,\mbox{and}\,\,
(k_3,\mu_3) \,\Bigg].
\end{eqnarray}
Here we defined 
$({\bm p} \cdot {\bm q})_\perp^{}=p^1q^1+p^2q^2$. These contributions
come from the HLL terms in Eqs. (\ref{two}) and (\ref{three}). 
The HLL contribution coming from
$\Delta_{HLL}^{\mu_{\parallel}\nu_{\parallel}\lambda_{\parallel}}$ in 
Eq.~(\ref{one}) is of a higher order in $1/|eB|$
and therefore is neglected here.
\footnote
{Note that the vertex for the initial non-smeared fields $A_\mu$ 
is given by 
$e^{-\frac{{\bm k}_{1\perp}^2+{\bm k}_{2\perp}^2+{\bm k}_{3\perp}^2}{4|eB|}}
T^{\mu_1\mu_2\mu_3}$. Thus, for these fields, an exponentially damping 
form-factor
occurs not in the propagator but in their vertices (compare with 
a discussion of this feature in Ref. \cite{GM}).}

Using Eqs.~(\ref{divergence}), (\ref{a}), and (\ref{b}),
one can easily check that 
the 3-point vertex $T^{\mu_1\mu_2\mu_3}$ is transverse.
The cancellation occurs between $T^{\mu_1\mu_2\mu_3}_{LLL}$ and 
$T^{\mu_1\mu_2\mu_3}_{HLL1}$.
As to the term $T^{\mu_1\mu_2\mu_3}_{HLL2}$,
it is transverse itself. 

Is there a kinematic region in which the LLL contribution 
is dominant? The answer to this question is ``yes". It is
the region with momenta
${\bm k}_{i\perp}^2 \gg |k_{i\parallel}^2|$. In this case,
the leading terms in the expansion of the LLL and HLL vertices
in powers of $k_{i\parallel}$ are:
\begin{equation}
  T^{\mu_1\mu_2\mu_3}_{LLL}(k_1,k_2,k_3) = 
  -\frac{2e\tilde{\alpha}}{3\pi}\frac{|eB|}{m^2} 
   \sin\left(\frac{1}{2}\theta_{ab}k_{1\,\perp}^a k_{2\,\perp}^b\right)
   \Bigg[\,(k_2-k_3)_\parallel^{\mu_1} g_\parallel^{\mu_2\mu_3}
          +(k_3-k_1)_\parallel^{\mu_2} g_\parallel^{\mu_3\mu_1}
          +(k_1-k_2)_\parallel^{\mu_3} g_\parallel^{\mu_1\mu_2}\,\Bigg], 
\end{equation}
\begin{eqnarray}
  T^{\mu_1\mu_2\mu_3}_{HLL1}(k_1,k_2,k_3) &=& 
  -\frac{2e\tilde{\alpha}}{3\pi}\frac{|eB|}{m^2} 
  \sin\left(\frac{1}{2}\theta_{ab}k_{1\,\perp}^a k_{2\,\perp}^b\right)
\Bigg[\,\dfrac{-k_{1\perp}^{\mu_1}}{{\bm k}_{1\perp}^2}\,
        \left\{\,(k_{2\parallel}^2-k_{3\parallel}^2)g_\parallel^{\mu_2\mu_3}
              - k_{2\parallel}^{\mu_2}k_{2\parallel}^{\mu_3} 
              + k_{3\parallel}^{\mu_2}k_{3\parallel}^{\mu_3}
        \,\right\} \nonumber \\ 
&& \hspace*{2.5cm}
      + \frac{k_{1\perp}^{\mu_1}k_{2\perp}^{\mu_2}
             -k_{1\perp}^{\mu_2}k_{2\perp}^{\mu_1}
            -({\bm k}_1 \cdot {\bm k}_2)_\perp^{} g_{\,\perp}^{\mu_1\mu_2}}
             {{\bm k}_{1\perp}^2\,{\bm k}_{2\perp}^2}\,
        \left\{\,k_{3\parallel}^2 k_{2\parallel}^{\mu_3}
       -(k_2 \cdot k_3)_\parallel k_{3\parallel}^{\mu_3}\,\right\} \nonumber \\
&& \hspace*{4cm} \qquad + \,\, 
\mbox{permutations of}\,\, (k_1,\mu_1),\, (k_2,\mu_2),\,\,\mbox{and}\,\,
(k_3,\mu_3) \,\Bigg],
\end{eqnarray}
\begin{eqnarray}
  T^{\mu_1\mu_2\mu_3}_{HLL2}(k_1,k_2,k_3) &=& 
  \frac{2e\tilde{\alpha}}{3\pi}\frac{eB}{m^2} 
\Bigg[\,\left\{\,\exp\left(\,-\frac{({\bm k}_1\cdot{\bm k}_2)_\perp^{}}
                                   {2|eB|}\,\right)
        -\cos\left(\frac{1}{2}\theta_{ab}k_{1\,\perp}^a k_{2\,\perp}^b\right)
        \,\right\} \nonumber \\ && \qquad \qquad \times
\left(\,\frac{k_{1\perp}^{\mu_2}\epsilon_\perp^{ab}k_{1\,\perp}^a k_{2\,\perp}^b
              +({\bm k}_1 \cdot {\bm k}_2)_\perp^{} \epsilon_\perp^{\mu_2 b}
               k_{1\,\perp}^b}
              {{\bm k}_{1\perp}^2\,{\bm k}_{2\perp}^2}\,\, 
          \left\{\,k_{3\parallel}^2 g_\parallel^{\mu_3\mu_1} - 
                   k_{3\parallel}^{\mu_3}k_{3\parallel}^{\mu_1}\,\right\}
\right. \nonumber \\ && \qquad \qquad \qquad 
        +\frac{k_{2\perp}^{\mu_1}\epsilon_\perp^{ab}k_{2\,\perp}^a k_{1\,\perp}^b
              +({\bm k}_2 \cdot {\bm k}_1)_\perp^{}
              \epsilon_\perp^{\mu_1 b}k_{2\,\perp}^b} 
              {{\bm k}_{1\perp}^2\,{\bm k}_{2\perp}^2}\,\, 
          \left\{\,k_{3\parallel}^2 g_\parallel^{\mu_2\mu_3} - 
                   k_{3\parallel}^{\mu_2}k_{3\parallel}^{\mu_3}\,\right\}
\nonumber \\ && \qquad \qquad \qquad \left.
        +\frac{g_{\,\perp}^{\mu_1\mu_2}\epsilon_\perp^{ab}k_{1\,\perp}^a k_{2\,\perp}^b
              +\epsilon_\perp^{\mu_1 \mu_2}({\bm k}_1 \cdot {\bm k}_2)_\perp^{}}
              {{\bm k}_{1\perp}^2\,{\bm k}_{2\perp}^2}\,\,\,
        \left\{\,k_{3\parallel}^2 k_{2\parallel}^{\mu_3}
       -(k_2 \cdot k_3)_\parallel k_{3\parallel}^{\mu_3}\,\right\}\,\right)
\nonumber \\ && \hspace*{4cm} \qquad + \,\, 
\mbox{permutations of}\,\, (k_1,\mu_1),\, (k_2,\mu_2),\,\,\mbox{and}\,\,
(k_3,\mu_3) \,\Bigg].
\end{eqnarray}
It is clear from these expressions that in that region the LLL 
contribution 
dominates indeed.
\footnote 
{Although the vertex $T^{\mu_1\mu_2\mu_3}_{HLL1}$ is subdominant
in this region, it is crucial for the restoration of the Ward
identity. It is because while in the Ward identity there are terms in
the HLL vertex $T^{\mu_1\mu_2\mu_3}_{HLL1}$ which are  
multiplied by a large transverse momentum
$k_{\perp}$, the LLL vertex $T^{\mu_1\mu_2\mu_3}_{LLL}$ is multiplied 
only by a small longitudinal
momentum $k_{\parallel}$.} 
This result is quite noticeable.
The point is that as was shown in
Ref. \cite{QED1}, the region with momenta ${\bm k}_{i\perp}^2 \gg 
|k_{i\parallel}^2|$ yields the dominant contribution in the
Schwinger-Dyson equation for the dynamical fermion mass in QED in a 
strong magnetic field. Therefore the LLL approximation is
reliable in that problem.


Nondecoupling of (heavy) HLL in the infrared region is quite
unexpected phenomenon. 
What physics underlines it? We believe 
that this phenomenon reflects the important role of a boundary dynamics  
at spatial infinity in this problem. The point is that the HLL are
not only heavy states but their transverse size grows without limit with 
their gap $\sqrt{m^2  + 2|eB|n}$ as $n \to \infty$. This happens because
the transverse dynamics in the Landau problem is oscillator-like one.
Since in order to cancel the LLL anomaly
one should consider the contribution of {\it all} higher Landau levels,
it implies that the role of the boundary dynamics at the 
transverse spatial infinity (corresponding to $n \to \infty$) is
crucial for the restoration of the gauge symmetry. In this respect 
this phenomenon is similar to that of edge states
in the quantum Hall effect: the edge states
are created by the boundary dynamics and also restore the gauge
invariance \cite{Hall}. 
Both these phenomena reflect the importance of
a boundary dynamics in a strong magnetic
field. It would be interesting to examine whether similar
nondecoupling phenomena take place in noncommutative theories arising
in string theories in magnetic backgrounds \cite{DNS}.

\begin{acknowledgments}
Discussions with A. Buchel, V. Gusynin, and I. Shovkovy are acknowledged.
We are grateful for support from the Natural 
Sciences and Engineering Research Council of Canada.

\end{acknowledgments}

\end{document}